\documentclass[aps,prstab,reprint,groupedaddress]{revtex4-1}
\pdfoutput=1
\usepackage{tikz}
\usepackage{graphicx}
\usepackage{url}
\usepackage{fancyvrb}
\usepackage{color}
\usepackage{tikz}
\usepackage{hyperref}
\usepackage{todonotes}

\usetikzlibrary{shapes,arrows,positioning,shapes.geometric}
\usepackage{charter}
\usepackage{siunitx}
\usepackage{amssymb, amsmath}
\usepackage{graphicx}
\usepackage{color}
\usepackage{afterpage}

\usepackage{struktex}

\usepackage{xspace}

\newcommand{\degree}{\ensuremath{^\circ}\xspace}

\newcommand{\htp}{\ensuremath{\mathrm{H}_2^+}\xspace}

\newcommand{\BE}[0]{\begin{equation}}
\newcommand{\EE}[0]{\end{equation}}
\newcommand{\BEA}[0]{\begin{eqnarray}}
\newcommand{\EEA}[0]{\end{eqnarray}}

\newcommand{\nuebar}{\ensuremath{\bar{\nu}_e}\xspace}

\mathchardef\mhyphen="2D

\newcommand{\figref}[1]{Figure~\ref{#1}}
\newcommand{\tabref}[1]{Table~\ref{#1}}

\newcommand{\opal}{\textsc{OPAL}\xspace}
\newcommand{\opalt}{\textsc{OPAL-t}\xspace}
\newcommand{\opalcycl}{\textsc{OPAL-cycl}\xspace}

\newcommand{\DD}{DAE$\delta$ALUS\xspace}
\newcommand{\trilinos}{\texttt{Trilinos}}

\newcommand {\bs}[1]{\mathbf #1}
\newcommand {\RM}[1]{\mathrm{#1}}

\begin{document}


\title{Realistic Injection Simulations of a Cyclotron Spiral Inflector using OPAL}


\author{Daniel Winklehner}
\email[]{winklehn@mit.edu}
\altaffiliation{Massachusetts Institute of Technology, 77 Massachusetts Ave, Cambridge, MA, USA}
\author{Andreas Adelmann}
\author{Achim Gsell}

\affiliation{Paul Scherrer Institut, 5232 Villigen PSI, Switzerland}

\author{Tulin Kaman}

\affiliation{Institute of Mathematics, University of Zurich, CH-8057, Switzerland}

\author{Daniela Campo}

\affiliation{INFN - INFN - Laboratori Nazionali di Legnaro, 35020 Legnaro, Italy}


\date{\today}

\begin{abstract}
We present an upgrade to the particle-in-cell ion beam simulation code OPAL
that enables us to run highly realistic simulations of the spiral inflector
system of a compact cyclotron. This upgrade includes a new geometry class and
field solver that can handle the complicated boundary conditions posed by
the electrode system in the central region of the cyclotron both in terms of
particle termination, and calculation of self-fields. 
Results are benchmarked against the analytical solution of a coasting beam.
As a practical example, the spiral inflector and the first revolution 
in a 1 MeV/amu test cyclotron, located at Best Cyclotron Systems, Inc., 
are modeled and compared to the simulation results. We find that OPAL
can now handle arbitrary boundary geometries with relative ease. 
Comparison of simulated injection efficiencies, and beam shape compare 
well with measured efficiencies and a preliminary measurement of the beam 
distribution after injection.
\end{abstract}


\maketitle


\section{Introduction \label{sec:intro}}
\opal \cite{adelmann:opal} is a particle-in-cell (PIC) code developed for the 
simulation of particle accelerators. It is highly parallel and comes in two 
distinct flavors: \opalcycl (specific to cyclotrons and rings) and \opalt 
(general purpose). 
For the presented application, we focus on 
\opalcycl\, which has been used very successfully to simulate existing 
high intensity cyclotrons like the PSI Injector II \cite{PhysRevSTAB.13.064201},
and PSI Ring Cyclotron \cite{PhysRevSTAB.13.064201, bi:cyclotron_sim}
as well as to design new cyclotrons like CYCIAE \cite{zhang:cyciae},
\DD \cite{abs:daedalus, aberle:daedalus}, 
and IsoDAR \cite{adelmann:isodar, bungau:isodar}.

\begin{figure}[!t]
\centering
	\includegraphics[width=0.9\columnwidth]
	                {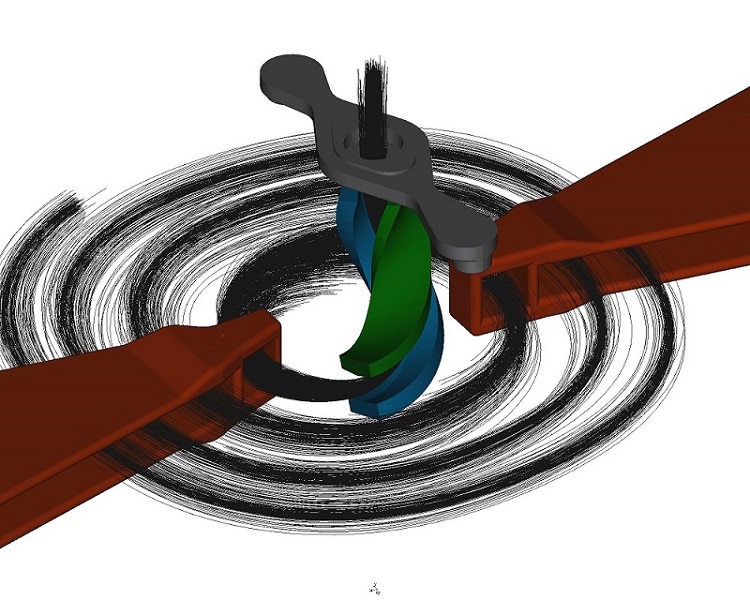}
	\caption{Schematic of a spiral inflector with particle trajectories from
			 an \opal simulation. The beam enters axially (from the top) 
			 through an aperture (grey) and is
			 bent into the mid-plane by a combination of the electrostatic field
			 generated by the spiral electrodes (green and blue, voltage 
			 typically symmetric at $+\mathrm{V}_\textrm{spiral}$ and 
			 $-\mathrm{V}_\textrm{spiral}$) and the cyclotron's main magnetic 
			 field. Then it is accelerated 
			 by the two Dees (copper, Dummy-Dees not shown). Color online.}
	\label{fig:spiral_schematic}
\end{figure}

However, one piece has been missing so far: the axial injection using a spiral
inflector or an electrostatic mirror. Both are electrostatic devices that bend 
the beam from the axial direction into the mid-plane of the cyclotron where
it is subsequently accelerated. A schematic view of this type of injection 
for a spiral inflector is shown in \figref{fig:spiral_schematic}.

In order to enable \opalcycl to track particles through a complicated set 
of electrodes like a spiral inflector, the following additions have been made:
\begin{itemize}
\item The Smooth Aggregation Algebraic Multi Grid (SAAMG) solver
\cite{adelmann:poisson} has been extended to include arbitrarily shaped boundaries.
\item A geometry class has been implemented that provides boundary conditions 
to the field solver and handles particle termination should their trajectories
intersect with the electrodes.
\item A number of improvements have been made to the internal coordinate 
transformations and the handling of beam rotations in \opalcycl in order
to accommodate the injection off-mid-plane.
\end{itemize}

These additions will be discussed in Section \ref{sec:opal} and references 
therein. At the end of this section, a benchmark against the analytical solution
of a coasting beam in a grounded beam pipe with variable axial offset 
will be presented.

A specific example of the usefulness of realistic spiral inflector simulations 
is the ongoing R{\&}D effort for the \DD \cite{abs:daedalus, aberle:daedalus}, 
and IsoDAR \cite{adelmann:isodar, bungau:isodar} experiments (described briefly
in Section \ref{sec:BCS}). In both cases, a very high intensity beam 
($\approx10-30$ mA average) needs to be injected, which is higher than current 
state-of-the-art cyclotrons have demonstrated. Part of the R{\&}D for these 
projects was an experiment in collaboration with Best Cyclotron Systems, Inc. (BCS)
in Vancouver, Canada to test a flat-field ECR ion source, transport through
the Low Energy Beam Transport System (LEBT) and finally injection into a small
test cyclotron through a spiral inflector.
The results of this campaign are described in much detail in \cite{winklehner:bcs_tests} and
the most important points will be reiterated in Section \ref{sec:BCS},
before OPAL simulation results are bench-marked against the experimental results.

\section{The particle-in-cell code OPAL\label{sec:opal}}
\label{sec:amodel}
For this discussion we briefly introduce \opalcycl\ \cite{PhysRevSTAB.13.064201}, one of the four flavours of \opal. 

\subsection{Governing equation}
The collision between particles can be neglected in the cyclotron under consideration, because the typical bunch densities are low.
The general equations of motion of charged particles in electromagnetic fields can be expressed in time domain by
\begin{equation}\label{eq:motion}
  \frac{d\bs{p}(t)}{dt}  = q\left(c\mbox{\boldmath$\beta$}\times \bs{B} + \bs{E}\right), \nonumber \\
\end{equation}
where $m_0, q,\gamma$ are rest mass, charge and the relativistic factor. With $\bs{p}=m_0 c \gamma \mbox{\boldmath$\beta$}$ we denote the momentum of a particle, 
$c$ is the speed of light, and $\mbox{\boldmath$\beta$}=(\beta_x, \beta_y, \beta_z)$ is the normalized velocity vector. The time ($t$) and position ($\bs{x}$) dependent electric and magnetic vector fields are
written in abbreviated form as $\bs{E}$ and $\bs{B}$.

If $\bs{p}$ is normalized by $m_0c$, 
Eq.\,(\ref{eq:motion}) can be rewritten component wise in Cartesian coordinates as 
\begin{eqnarray}
  \frac{dp_x}{dt} & = & \frac{q}{m_0c}E_x + \frac{q}{\gamma m_0}(p_y B_z - p_z B_y),    \nonumber \\
  \frac{dp_y}{dt} & = & \frac{q}{m_0c}E_y + \frac{q}{\gamma m_0}(p_z B_x - p_x B_z),   \\
  \frac{dp_z}{dt} & = & \frac{q}{m_0c}E_z + \frac{q}{\gamma m_0}(p_x B_y - p_y B_x).    \nonumber 
\label{eq:motion2}
\end{eqnarray}

The evolution of the beam's distribution function 
\[
f(\bs {x},c\mbox{\boldmath$\beta$},t): (\Re^M \times \Re^M \times \Re) \rightarrow \Re
\]
can be expressed by a collisionless Vlasov equation:
\begin{equation}\label{eq:Vlasov}
  \frac{df}{dt}=\partial_t f + c\mbox{\boldmath$\beta$} \cdot \nabla_x f +q(\bs{E}+ c\mbox{\boldmath$\beta$}\times\bs{B})\cdot \nabla_{c\mbox{\boldmath$\beta$}} f  =  0.
\end{equation}
Here we have assumed that $M$ particles of the same species are within the beam.\
In this particular case, $\bs{E}$ and $\bs{B}$ include both external applied fields and space charge fields, all other fields are neglected.
\begin{eqnarray}\label{eq:Allfield}
  \bs{E} & = & \bs{E_{\RM{ext}}}+\bs{E_{\RM{sc}}}, \nonumber\\    
  \bs{B} & = & \bs{B_{\RM{ext}}}+\bs{B_{\RM{sc}}}.
\end{eqnarray}


\subsection{Bunch rotations}
In \opalcycl, the coordinate system is different from \opalt:
The x and y coordinates are the horizontal coordinates (the cyclotron mid-plane) and
z is the vertical coordinate. Internally, both Cartesian (x, y, z) and cylindrical 
(r, $\Theta$, z) coordinate systems are used.
For simplicity, in the past, the injection of bunches in \opalcycl had to happen on 
the cyclotron mid-plane with the option to include an offset in z direction within 
the particle distribution itself.
The global coordinates of the beam centroid were thus restricted to r and $\Theta$.
In order to accommodate the injection of a bunch far away from the mid-plane and a 
mean momentum aligned with the z-axis, the handling of rotations in \opalcycl
was updated from 2D rotations in the mid-plane to arbitrary rotations/translations
in three-dimensional (3D) space.
Quaternions \cite{hamilton:quaternions} were chosen to avoid gimbal-lock
\cite{mitchell:quaternions} and a set of rotation functions was implemented.
These are now used throughout \opalcycl. Outside of the new \verb|SPIRAL| mode,
quaternions are also used to align the mean momentum of the bunch with 
the y axis before solving for self-fields in order to simplify the inclusion
of relativistic effects in the calculation of self-fields. 
In the \verb|SPIRAL| mode, no relativistic effects are taken into account due to
the low injection energy (typically $<100$ keV). 

\subsection{Fieldsolver} \label{ssec:fieldsolver}

The space charge fields can be obtained
by a quasi-static approximation. In this approach, the relative motion of the particles is non-relativistic in the beam rest frame, so the self-induced magnetic field is practically absent and the electric field can be computed by solving Poisson's equation
\begin{equation}\label{eq:Poisson}
  \nabla^{2} \phi(\bs{x}) = - \frac{\rho(\bs{x})}{\varepsilon_0},
\end{equation}
where $\phi$ and $\rho$ are the electrostatic potential and the spatial charge density in the beam rest frame. The electric field can then be calculated by
\begin{equation}\label{eq:Efield}
  \bs{E}_{\RM{sc}}=-\nabla\phi,
\end{equation}
and back transformed to yield both the electric and the magnetic fields, in the lab frame, required in Eq.\,(\ref{eq:Allfield}) by means of a Lorentz transformation.
As mentioned in the previous section, this step is omitted in the \verb|SPIRAL| mode 
of \opalcycl by setting $\gamma = 1$. 

A parallel 3D Poisson solver for the electrostatic potential 
computation in a round beam pipe geometry with open-end boundary conditions 
is presented in ~\cite{adelmann:poisson}. 

In this section, we present a method for solving the space charge Poisson 
problem~\eqref{eq:poisson} in much more complex geometries of particle accelerators. 
The problem discretization takes into account the complex geometries of beam pipe elements 
and the space charge forces are computed within the geometry. This assures that the space 
charge components are taken properly into account in any type of geometry.
\begin{equation} \label{eq:poisson}
  \begin{aligned}
    -\Delta \phi &= \frac{\rho}{\epsilon_0}\ \text{in}\ \Omega, \\
    \phi &= g \equiv 0\ \text{on}\ \partial\Omega \\
  \end{aligned}
\end{equation}
where $\Omega$ represents the computational domain, and $\partial\Omega$ the boundary surface of the geometry.\
Our approach solving the Poisson problem in complex geometries includes the unstructured 
and structured meshes. Unstructured meshes are used to define the complex geometry because 
they are easy to conform a block to a complicated shape, and structured meshes are used 
for the finite-difference discretization method.  

\subsubsection{Spatial discretization}\label{ssec:discretization}

To solve the Poisson problem~\eqref{eq:poisson} on $\Omega$, we
use a cell-centered discretization for Laplacian.  A  grid point is 
called interior if all its neighbors are in $\Omega$, or near-boundary point otherwise.
The discrete Laplacian with mesh spacing $h$ on interior points 
is defined as the regular seven-point finite difference 
approximation of~$-\Delta$.
\begin{equation}  \label{eq:potentialdisc}
 L{\hat\phi}(\mathbf{x})= \sum_{i=1}^3 \frac{-{\hat
      \phi}(\mathbf{x}-h\mathbf{e}_i) 
    +2{\hat \phi}(\mathbf{x})-{\hat \phi}(\mathbf{x}+h\mathbf{e}_i)}{h^2},
\end{equation}
where $\mathbf{e}_i$ is the $i$th coordinate vector~\cite{fowa:60}.

\begin{figure}[htb]
  \centering
  \includegraphics[width=0.3\textwidth]{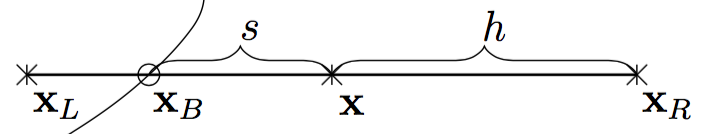}
  \caption{$\mathbf{x}_L=\mathbf{x}-h\mathbf{e}_i$, 
$\mathbf{x}_B=\mathbf{x}-sh\mathbf{e}_i$,
$\mathbf{x}$ and $\mathbf{x}_R=\mathbf{x}+h\mathbf{e}_i$ are 
the point outside of $\Omega$, boundary point, near-boundary point
and interior point.
}
\label{fig:extrapolation}
\end{figure}

At near-boundary point, $\mathbf{x}$ in \figref{fig:extrapolation} 
where not all its neighbors are in the domain, 
we use an approximation for Laplacian and
the value of $\hat{\phi}$ at near-boundary point is obtained by one of the 
extrapolation method described in \cite{adelmann:poisson}. 
The approximation is based on linear extrapolation of near-boundary points 
($0 < s \leq 1$):
\begin{equation}
L{\hat\phi}(\mathbf{x})= \sum_{i=1}^3 
    -\frac{1}{sh}{\hat \phi}(\mathbf{x}_L) 
    +\frac{1}{h}\left(\frac{1}{s}+\frac{1}{h}\right){\hat \phi}(\mathbf{x})
    -\frac{1}{h^2}{\hat \phi}(\mathbf{x}_R),
\end{equation}

The value of ${\hat\phi}$ at $\mathbf{x}_L$ is defined through one 
of the extrapolation method in Eq.~\eqref{eq:extrapolation}. 
\begin{equation} \label{eq:extrapolation}
  \begin{aligned}
  \rm{Constant:} \  {\hat\phi}(\mathbf{x}_L) &= g(\mathbf{x}_B)  \\
  \rm{Linear:} \  {\hat\phi}(\mathbf{x}_L) &= \left ( 1- \frac{1}{s}\right){\hat\phi}(\mathbf{x}) 
    + \frac{1}{s} g(\mathbf{x}_B)  \\
  \end{aligned}
\end{equation}

\subsubsection{Implementation}\label{ssec:impl}

The query class, an interface to search for the points inside of the irregular domain 
is implemented. Once the points inside the irregular domain are detected,
their intersection values in six different directions are stored in containers.    
The coordinates values are mapped into its intersection values to be used as 
a fast look-up table. The distances between the near-boundary point and its intersection 
values are used for the linear extrapolation.    

The finite difference approximation of the Poisson problem requires solving a system 
of linear equations to compute the electrostatic potential. The resulting linear system is 
solved using the preconditioned conjugate gradient algorithm complemented by an 
algebraic multigrid preconditioner using the \trilinos\ framework~\cite{TrilinosArticle}.
\trilinos\ is a collection of software packages that support parallel linear algebra
computations on distributed memory architectures, in particular the solution of
linear systems of equations.  \texttt{Epetra} provides
the data structures that are needed in the linear algebra libraries.
\texttt{Amesos}, \texttt{AztecOO}, and \texttt{Belos} are packages providing
direct and iterative solvers.  \texttt{ML} is the multi-level package, that
constructs and applies the smoothed aggregation-based multigrid preconditioners.

\subsection{External fields}

\afterpage{\clearpage}
\begin{figure*}[t]
	\centering
	\begin{tabular}{cc}
		\includegraphics[height=240pt]{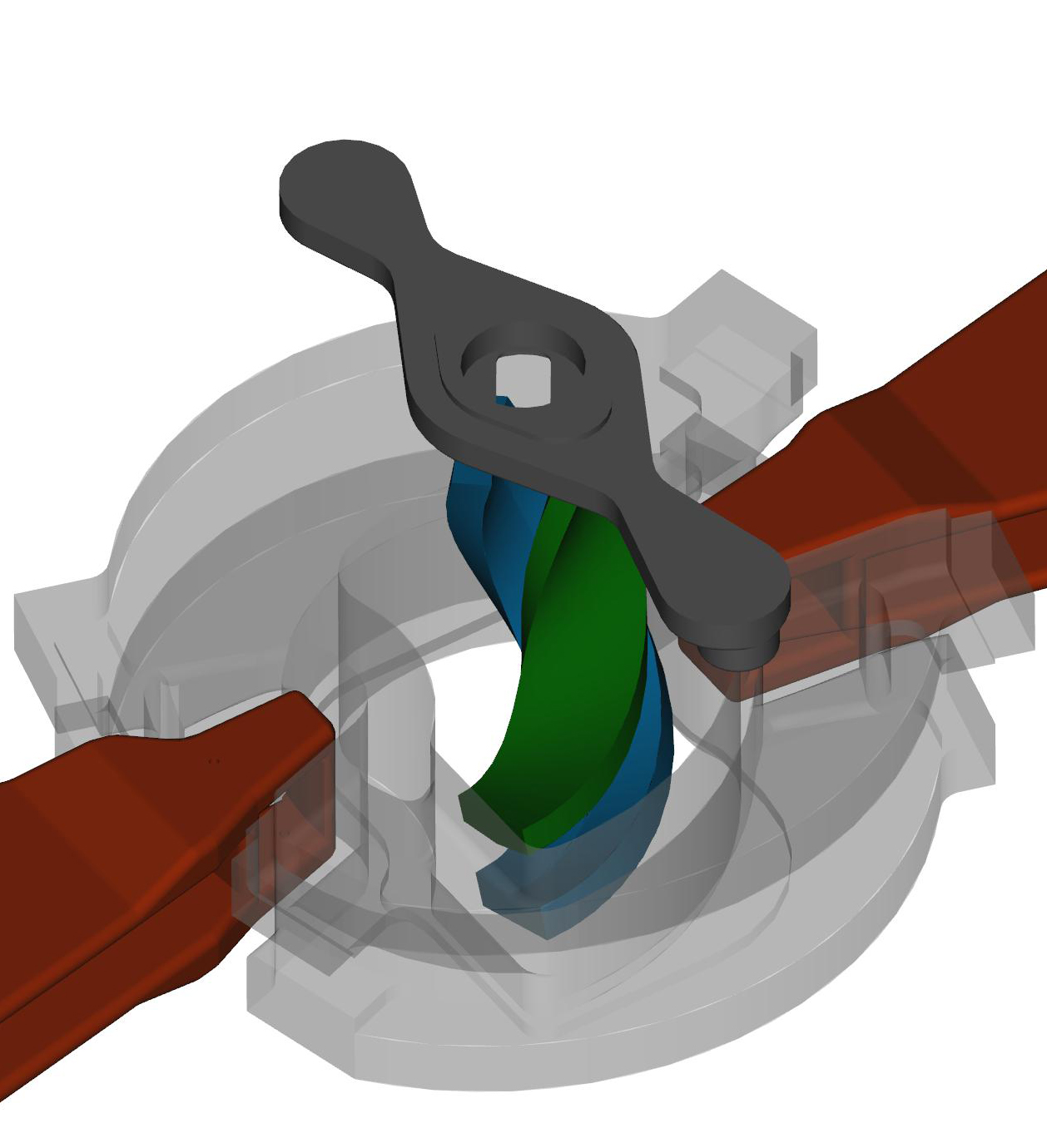} &
		\hspace*{40pt}\includegraphics[height=240pt]{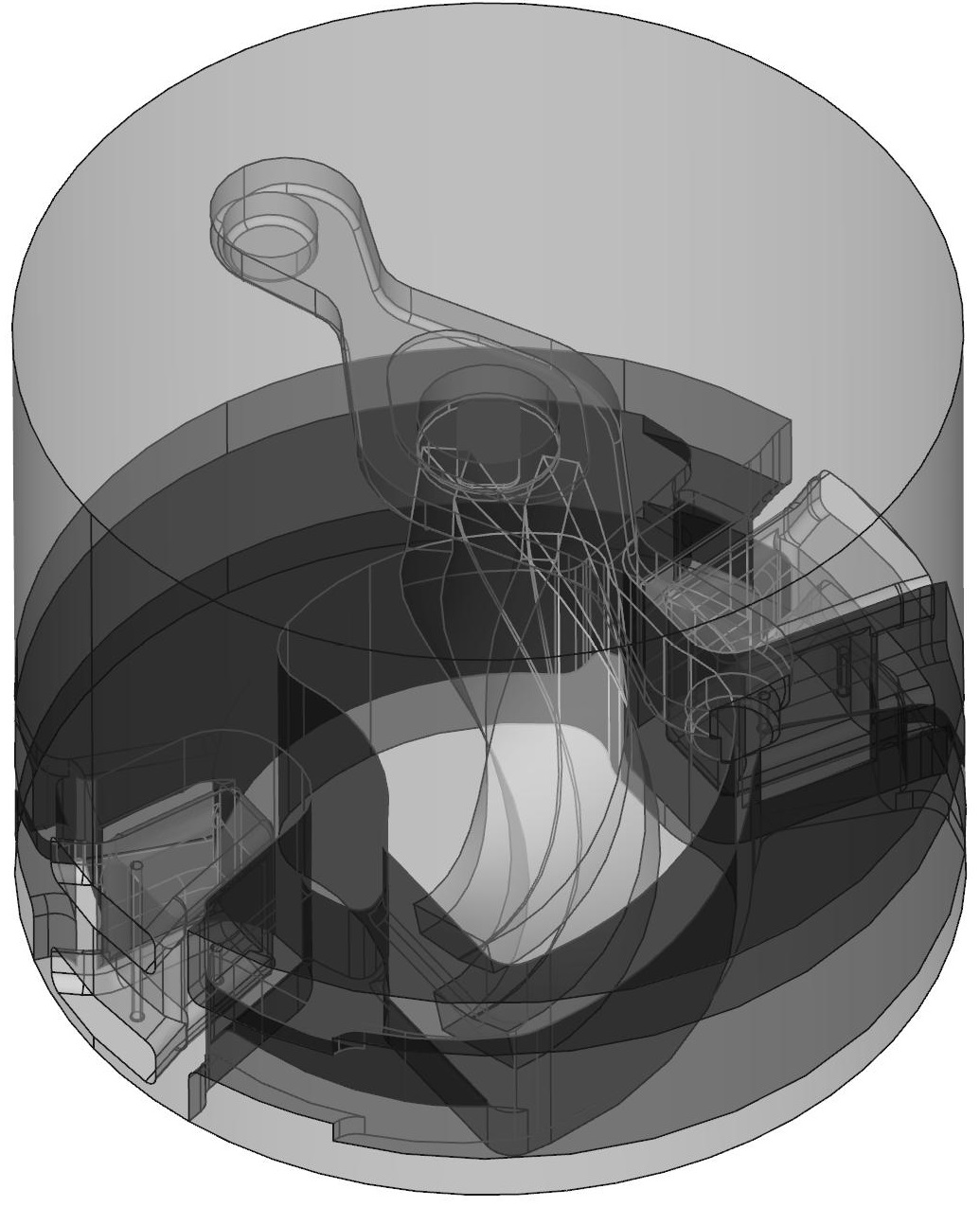} \\
		(a) CAD Model & 
		(b) Inverted \\
	    \includegraphics[height=240pt]{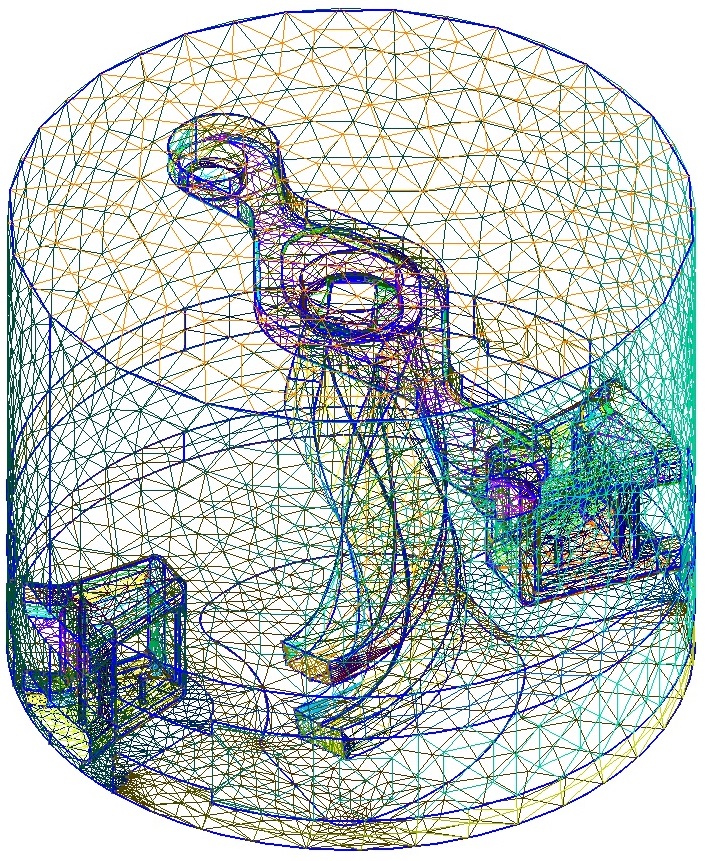} &
		\includegraphics[height=230pt]{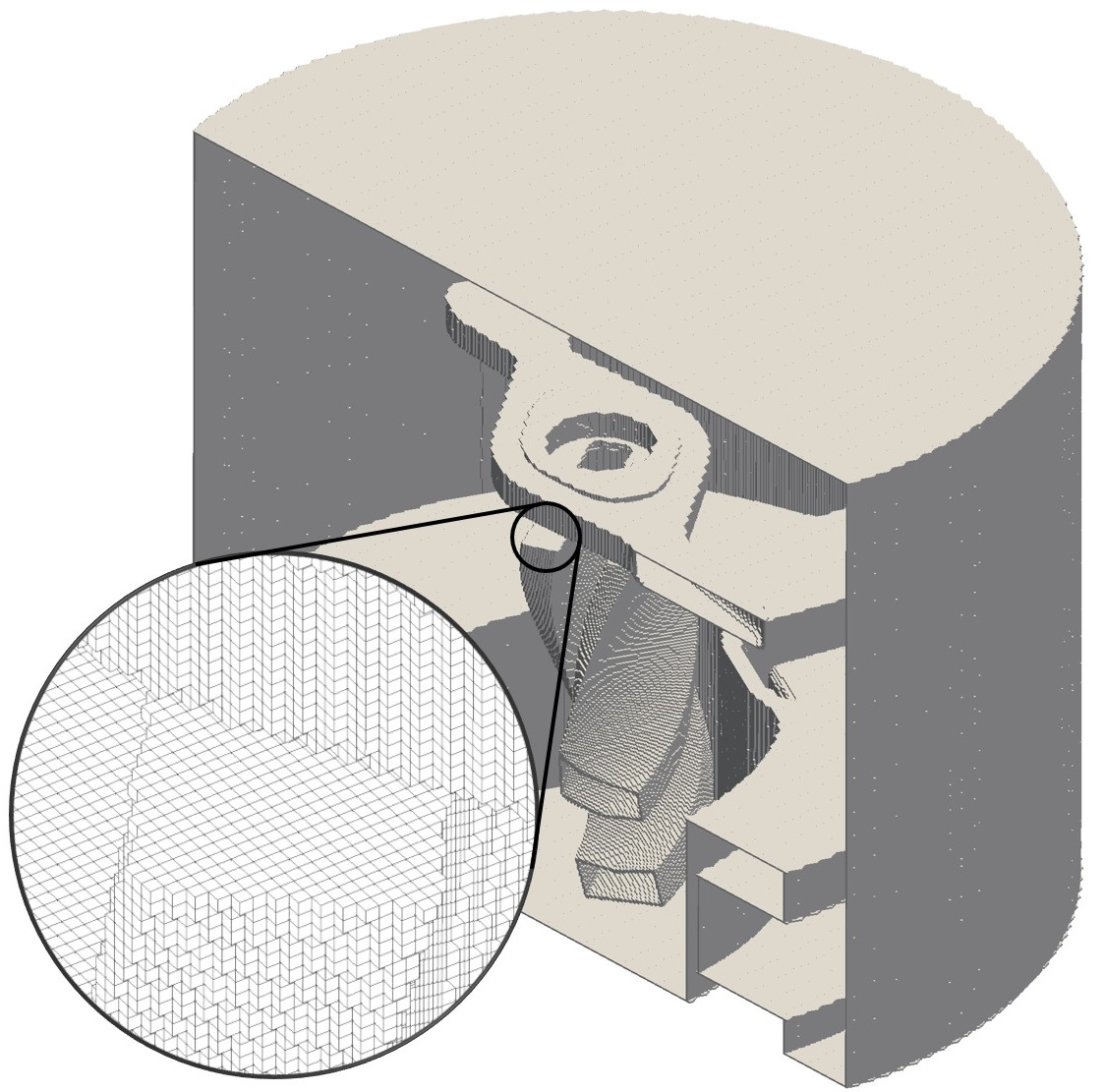} \\
		(c) Meshed & 
		(d) Voxels \\
	\end{tabular}
	\caption{The four stages of OPAL geometry preparation: 
			 (a) Initial CAD model of the electrodes in the system.
	         (b) An inverted solid is created by subtraction of all elements in (a) from
	             a ``master volume'' (in this case a cylindrical outer chamber).
	         (c) The inverted geometry is saved in stp format and a mesh is 
	             generated using GMSH \cite{geuzaine:gmsh}.
	         (d) During the initialization phase, \opal creates a voxel mesh to 
	             speed up the tests that have to be performed every time-step
	             (cf. text).}
    \label{fig:geometry}
\end{figure*}

With respect to the external magnetic field two possible situations can be considered: 
in the first situation, the real field map is available on the median plane of the existing cyclotron machine using measurement equipment. 
In most cases concerning cyclotrons, the vertical field, $B_z$, is measured on the median plane ($z=0$) only.
Since the magnetic field outside the median plane is required to compute trajectories with $z \neq 0$, the field needs to be expanded in the $Z$ direction. 
According to the approach given by Gordon and Taivassalo \cite{Gordon:2}, by using a magnetic potential and measured $B_z$ on the median plane
at the point $(r,\theta, z)$ in cylindrical polar coordinates, the 3$rd$ order field can be written as    
\begin{equation}
\vec{B}_{\RM{ext}} (r,\theta, z) = 
\begin{pmatrix}
z\frac{\partial B_z}{\partial r}-\frac{1}{6}z^3 C_r \\
\frac{z}{r}\frac{\partial B_z}{\partial\theta}-\frac{1}{6}\frac{z^3}{r}C_{\theta} \\
B_z-\frac{1}{2}z^2 C_z
\end{pmatrix}
\label{eq:Bfield}
\end{equation}

where $B_z\equiv B_z(r, \theta, 0)$ and  
\begin{eqnarray}\label{eq:Bcoeff}
  C_r & = & \frac{\partial^3B_z}{\partial r^3} + \frac{1}{r}\frac{\partial^2 B_z}{\partial r^2} - \frac{1}{r^2}\frac{\partial B_z}{\partial r} 
        + \frac{1}{r^2}\frac{\partial^3 B_z}{\partial r \partial \theta^2} - \frac{2}{r^3}\frac{\partial^2 B_z}{\partial \theta^2}, \nonumber  \\    
  C_{\theta} & = & \frac{1}{r}\frac{\partial^2 B_z}{\partial r \partial \theta} + \frac{\partial^3 B_z}{\partial r^2 \partial \theta}
        + \frac{1}{r^2}\frac{\partial^3 B_z}{\partial \theta^3},  \\
  C_z & = & \frac{1}{r}\frac{\partial B_z}{\partial r} + \frac{\partial^2 B_z}{\partial r^2} + \frac{1}{r^2}\frac{\partial^2 B_z}{\partial \theta^2}. \nonumber
\end{eqnarray}

All the partial differential coefficients are computed on the median 
plane data by interpolation, using Lagrange's 5-point formula.
In the other situation, the 3D field for the region of interest 
is calculated numerically by building a 3D model using commercial 
software during the design phase of a new cyclotron.
In this case the calculated field will be more accurate, 
especially at large distances from the median plane, i.e.
a full 3D field map can be calculated. 
For all calculations in this paper, we use the second method.
Fields (where applicable) are generated as 3D field maps 
with VectorFields OPERA \cite{opera:online}.
In case of RF fields, \opalcycl varies the field with a cosine function:
\begin{equation}
\label{ }
\vec{E}_{\RM{OPAL}}(t) = \vec{E}_{\RM{OPERA}} \cdot \cos(\omega_{\RM{rf}} t - \phi_S)
\end{equation}
where $t$ is the time of flight, $\omega_{\RM{rf}}$ the RF frequency,
 and $\phi_S$ the starting phase of the particle.
%
%
%

Finally, in this paper, both the external fields and self-fields 
are used to track particles during each time step
using a 4\textsuperscript{th} order Runge-Kutta (RK) integrator, 
in which the fields are evaluated four times in each time step. 
Space charge fields are assumed to be constant during one time step,
because their variation is typically much slower than that of 
external fields.

\subsection{Geometry}
For the simulation of precise beam dynamics, an exact modelling of the
accelerator geometry is essential. Usually a CAD model of the
accelerator or part of it is already available (\figref{fig:geometry}(a)). From these models we need to create the vacuum chamber or beam-pipe for specifying the boundary, i.e.\ the simulation domain ($\Omega$ in Eq. \ref{eq:poisson}). In case of the spiral inflector, we have to add a cylinder to limit the vacuum chamber (figure \figref{fig:geometry}(b)). This modified CAD model can be used to create a triangle mesh $T$ modeling the vacuum chamber of the inflector (\figref{fig:geometry}(c)).

Facilitating a meshed vacuum chamber or beam-pipe, \opal is able to model arbitrary
accelerator geometries and provides methods for
\begin{enumerate}
\item testing whether a particle will collide with the inner surface of
the geometry (boundary, $\partial\Omega$) in the next time step
\item computing the distance $d = | x_{0}-I |$ from a given point $x_{0}$ to the boundary intersection point $I$ with $\partial\Omega$ (c.f. \figref{fig:linetriag})
\item testing whether a given point $x_{0}$ is inside the geometry. Only points inside the geometry are in the computational domain $\Omega$.
\end{enumerate}

The geometry can consist of multiple parts. Each part must be modeled as a 
3D closed volume. The used methods are based on well 
known methods in computer graphics, especially ray 
tracing \cite{sunday:linetri}. 

\subsubsection{Initializing the geometry}
For testing whether a particle will collide with the boundary in then
next time step, we can run a line segment/triangle intersection test
for all triangles in the mesh. Even to be able to model simple structures,
triangle meshes with thousands of elements are required. Applying a brute force algorithm,
we have to run this test for all particles per
time-step, rendering the naive approach as not feasible due to
performance reasons.

In computer graphics this problem is efficiently solved by using voxel
meshes. A voxel is a volume pixel representing a value on a regular
3D grid. Voxel meshes are used to render and model 3D objects.

To reduce the number of required line segment -- triangle intersection
tests, a voxel mesh $V$ covering the triangle mesh $T$ is created during
initialization of \opal. In this step, all triangles are assigned to
their intersecting voxels. Whereby a triangle usually intersects with more
than one voxel.

For the line segment/triangle intersection tests we can now reduce the required tests to the triangles assigned to the voxels intersecting the line segment. The particle boundary collision test can be improved further by comparing the particle momentum and the inward pointing normal $\vec{n}$ of the triangles.

\begin{figure}[t]
\centering
\scalebox{0.85}{
\begin{tikzpicture}
\usetikzlibrary{arrows}
\draw [->,black,-latex] (-1.5,0) -- (1.5,0);
\draw [->,black,-latex] (-1.5,0) -- (-1.5,2);
\draw (1.5,0) -- (0.0,2);
\draw  [<-,black,latex-](0.0,2) -- (-1.5,0.0);
\node[below] (w) at (-0.25,0.5) {$t \in T$};
\node[above] (t0) at (-1.7,0.) {$\mathbf{t_0}$};
\node[right] (t1) at (1.5,0) {$\mathbf{t_1}$};
\node[right] (t1) at (3,2) {$\partial\Omega$};
\node[above] (t2) at (0.0,2.0) {$\mathbf{t_2}$};
\node[above] (n) at (-1.5,2) {$\vec{n}$};
\path[draw=black,fill=black] (0.0,2.0) circle (2pt);
\path[draw=black,fill=black] (1.5,0.0) circle (2pt);
\path[draw=black,fill=black] (-1.5,0.0) circle (2pt);
\draw (-3.5,-1) -- (2.5,-1);
\draw (2.5,-1) -- (5,3);
\draw (0,3) -- (5,3);
\draw (-3.5,-1) -- (0,3);
\draw (0.5,0.5) -- (1.6,5);
\path[draw=black] (0.5,0.5) circle (2pt);
\node[above] (I) at (0.3,0.5) {$\mathbf{I}$};
\node[right] (x0) at (1.6,5) {$\mathbf{x_{0}}$};
\draw[dashed] (0.5,0.5) -- (-0.1,-1.9);
\node[right] (x1) at (0.0,-1.9) {$\mathbf{x_1}$};
\node[right] (ri) at (1,2.5) {$d$};
\path[draw=black,fill=black] (-0.1,-1.9) circle (2pt);
\path[draw=black,fill=black] (1.6,5) circle (2pt);
\end{tikzpicture}}
\caption{The boundary $\partial\Omega$ is discretized by a set of triangles $T$. Shown is
the line $(x_{0},x_{1})$ triangle $t \in T$ intersection test, with intersection at $\mathbf{I} \in \partial\Omega$.}
    \label{fig:linetriag}
\end{figure}
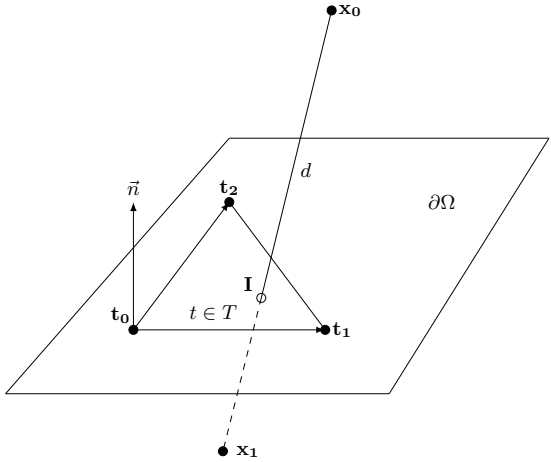

In the following, we use the following definitions:

\begin{description}
\item[$T$] represents the set of triangles in the triangulated mesh.

\item[$V$] represents the set of voxels modelling the voxelized triangle mesh.

\item[$L$] a closed line segment bounded by points $x_{0}$ and $x_{1}$ (c.f.\ figure \figref{fig:linetriag}).

\item[$R$] a ray defined by the starting point $x_{0}$ passing through $x_{1}$.

\item[$T_v \subset T$] represents the subset of triangles $t \in T$ which have intersections with $v \in V$.

\item[$V_L \subset V$] represents the subset of voxels $v \in V$ which intersections with the line segment $L$.

\item[$I_{t,L}$] represents an intersection point of a line segment $L$ with a triangle $t \in T$.

\item[$T_L$] represents the set of tuples $(t_L,I_{t,L})$ with $t_L \in T$ intersects with $L$.
\end{description}

\subsubsection{Basic ray/line-segment boundary intersection test}
In the first step, we have to compute $V_L$, which is the set of voxels in the voxel mesh $V$ which have intersections with the given line-segment or ray $L$. With 
$$T_v=\{t \in T\, |\, t\,\textrm{\it intersects with }v\}$$
we can compute a small subset of triangles
$$T_{V,L} = \{\cup T_v, \forall\, v \in V_L\}$$
which might have intersections with $L$. 
We have to run the ray/line-segment triangle test only for all triangles in $T_{V,L}$.

\begin{center}
\begin{struktogramm}(85, 55)[\textbf{basicIntersectionTest(L)}]
	\assign{$V_L = \{v \in V\, |\, v\, \textrm{\it intersects with } L\}$ }
	\assign{$T_{V,L} = \{t \in T\,| \,t\, \textrm{\it intersects with a voxel } v \in V_L\}$ }
	\assign{$T_L = \{\}$ // result}
	\while{$\textrm{\it for each}\, t \in T_{V,L}$}
		\ifthenelse{5}{1}{$t\, \textrm{\it intersects with } L?$}{$\textrm{\it yes}$}{$\textrm{\it no}$}
			\assign{$T_L\mathrel{{+}{=}}\{(t,I)\}$ // $I$: intersec. point}
		\change
		\ifend
	\whileend
	\assign{$\textrm{\it return } T_L$}
\end{struktogramm}
\end{center}

\subsubsection{Particle boundary collision test}
To test whether a particle will collide with the boundary we have to
test whether the line-segment given by the particle position $x_0$ at time step $s$
and the expected particle position $x_1$ at time $s+1$ intersect with the
boundary. The closed line segment $L$ given by $x_0$ and $x_1$ is used as input parameter for the boundary collision test.

\begin{center}
\begin{struktogramm}(85,55)[\textbf{boundaryCollisionTest(L)}]
	\assign{$\textrm{\it compute }T_L\, \textrm{{\it with} {\bf basicIntersectionTest()}} $}
	\assign{$I_L = \infty$ // intersection point to return}
	\assign{$t_L = -1$  // corresponding triangle}
	\while{$\textrm{\it for each } (t,I) \in T_L$}
		\ifthenelse{5}{1}{$\vert I - x_0 \vert < \vert I_L - x_0 \vert ?$}{$\textrm{\it yes}$}{$\textrm{\it no}$}
			\assign{$I_L = I$}
			\assign{$t_L = t$}
		\change
		\ifend
	\whileend
	\assign{$\textrm{\it return } (t_l,I_L)$}
\end{struktogramm}
\end{center}

In the collision test we use a slightly modified intersection test. If the particle moves away from a given triangle $t$, we do not have to run the line-segment triangle intersection test for $t$. Since we know the inward pointing normals $\vec{n}$ for all triangles, hence we can compute the dot product of the normal  $\vec{n}$, and the vector defined by the current particle position $x_{1}$ and the position in the next time step $x_1$. If
$$\vec{n} \cdot (x_1 - x_0) > 0$$
both vectors point in the same direction, so there cannot be a collision of the particle with the boundary.

\subsubsection{Compute distance from point to boundary}
To compute the distance from a point $P$ inside the geometry to the boundary in a given direction $v$, the same algorithm as for the particle boundary collision test can be used with the ray $L = (P,v)$.

\subsubsection{Inside test}
Test whether a given point $P$ is inside the geometry.

\begin{center}
\begin{struktogramm}(80,35)[\textbf{isInsideTest(P)}]
	\assign{$\textrm{\it select } R\, \textrm{\it outside geometry}$}
	\assign{$L = (R,P)$  // line-segment from $R$ to $P$}
	\assign{$\textrm{\it compute } T_L\, \textrm{{\it with} {\bf basicIntersectionTest()}}$}
	\ifthenelse{4}{4}{$\textrm{\it cardinality of } T_L\, \textrm{\it odd?}$}
	                 {$\textrm{\it yes}$}
	                 {$\textrm{\it no}$}
		\assign{$R\, \textrm{\it is inside}$}
	\change
		\assign{$R\, \textrm{\it is outside}$}
	\ifend
\end{struktogramm}
\end{center}

\subsection{Simple Test Case}

\begin{figure}[!t]
\centering
	\includegraphics[height=5.5cm]
	                {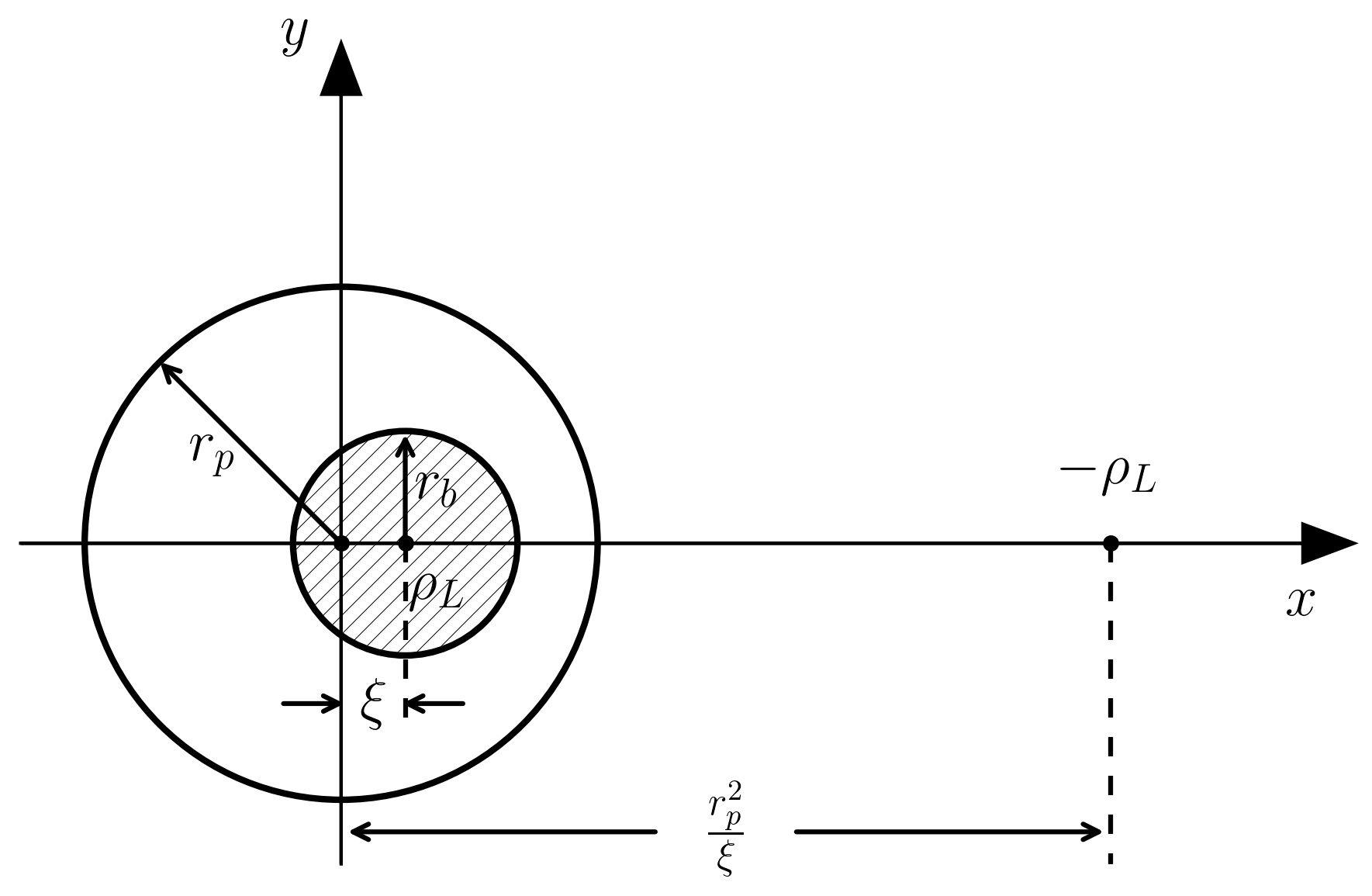}
	\caption{Cross-sectional view of beam off-centered
	         by $\xi$ along the x-axis. A mirror line charge
	         of $-\rho_L$ is also depicted at $\frac{r_p^2}{\xi}$.			 
	         Adapted from \cite[p229]{Reiser}.}
	\label{fig:offcenter}
\end{figure}

In order to test the proper functionality of the SAAMG
fieldsolver in combination with an external geometry file,
we compared the calculated fields and potentials of a
\verb|FLATTOP| distribution (uniformly populated cylinder)
inside a geometry generated according to the previous subsection
with the analytical solution of a uniformly charged cylindrical
beam inside a conducting pipe. 
The geometry file we used contained a 1 m long beam pipe with
0.1 m radius.
For a concentric beam, a regular Fast Fourier Transform (FFT) 
field solver is sufficient. However, if the beam is off-centered
by an amount $\xi$ (see \figref{fig:offcenter})
and especially, when it is close to the conducting walls of the beam
pipe, the electric field calculated by the FFT solver does no
longer reproduce reality. This is even more true for complicated 
geometries like a spiral inflector.
To compare the simulated results with the analytical solution 
presented below, all simulations were run in such a way 
that the bunch frequency $f_b$ was adjusted such that for a given 
bunch length $l_b$ and beam velocity $v_b$ the given beam current $I$ 
corresponded to the equivalent DC beam current (i.e. subsequent bunches 
are head-to-tail):
\BE
f_b = \frac{v_b}{l_b}
\EE
The bunch radius $r_b$ was chosen to be 0.01 m and the length $l_b$ 
to be 0.6 m so that in the center of the bunch to very good 
approximation the conditions of an infinitely long beam hold. 

\begin{figure}[!t]
\centering
	\includegraphics[height=5.5cm]
	                {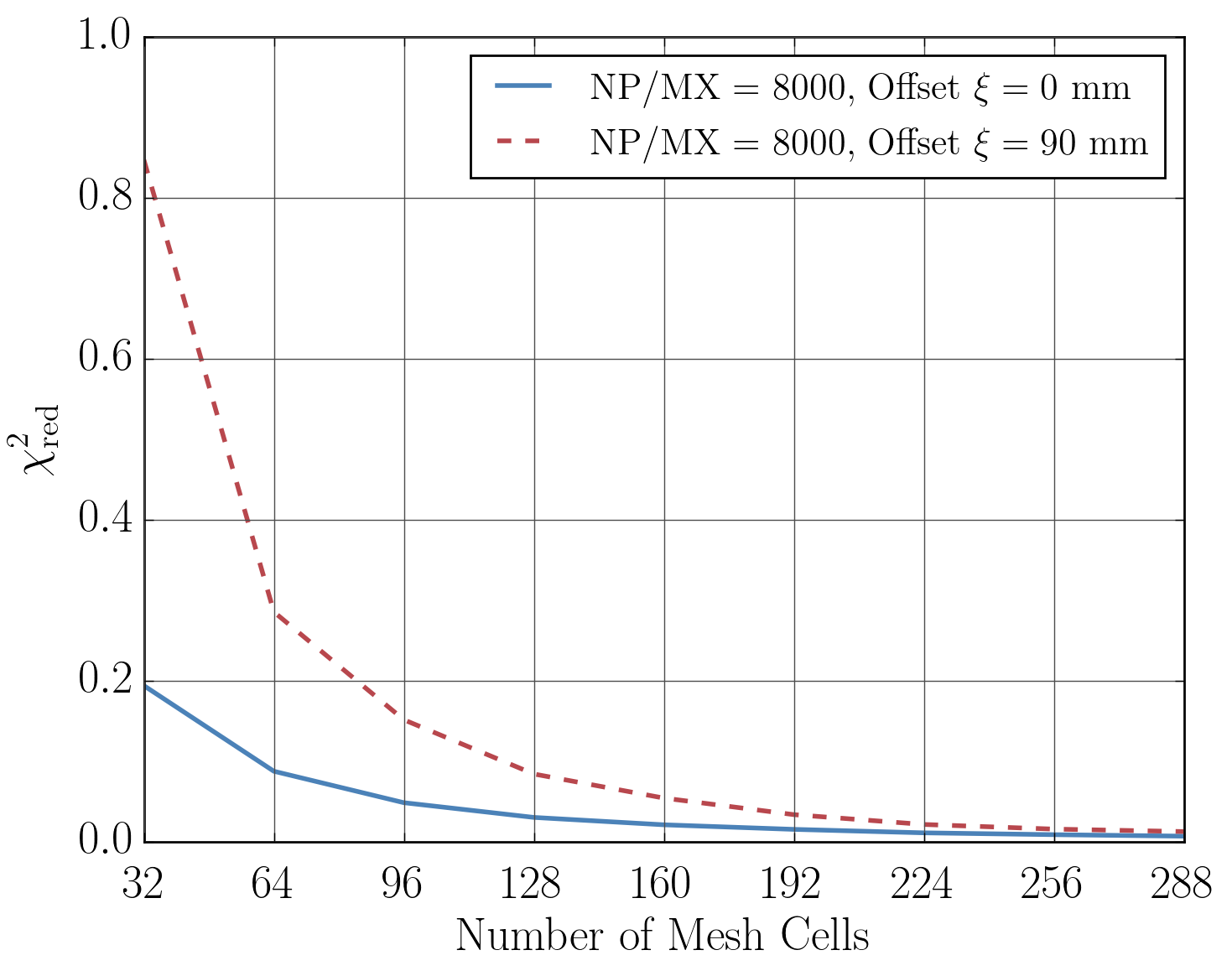}
	\caption{$\chi^2_{\mathrm{red}}$ of the calculated $\Phi$
	         compared to the simulated values. It can be seen that the results
	         are slightly worse for a beam close to the beam pipe.
	         Also, for number of mesh cells in x-direction $> 256$, only marginal
	         improvement can be seen. These simulations were performed with 
	         a total number of particles $\mathrm{NP} = 8000\cdot\mathrm{MX}$
	         where MX denotes the number of mesh cells in x-direction
	         (the independent variable in the plot).}
	\label{fig:chisquares}
\end{figure}

\begin{figure*}[t!]
\centering
	\includegraphics[width=0.8\textwidth]
	                {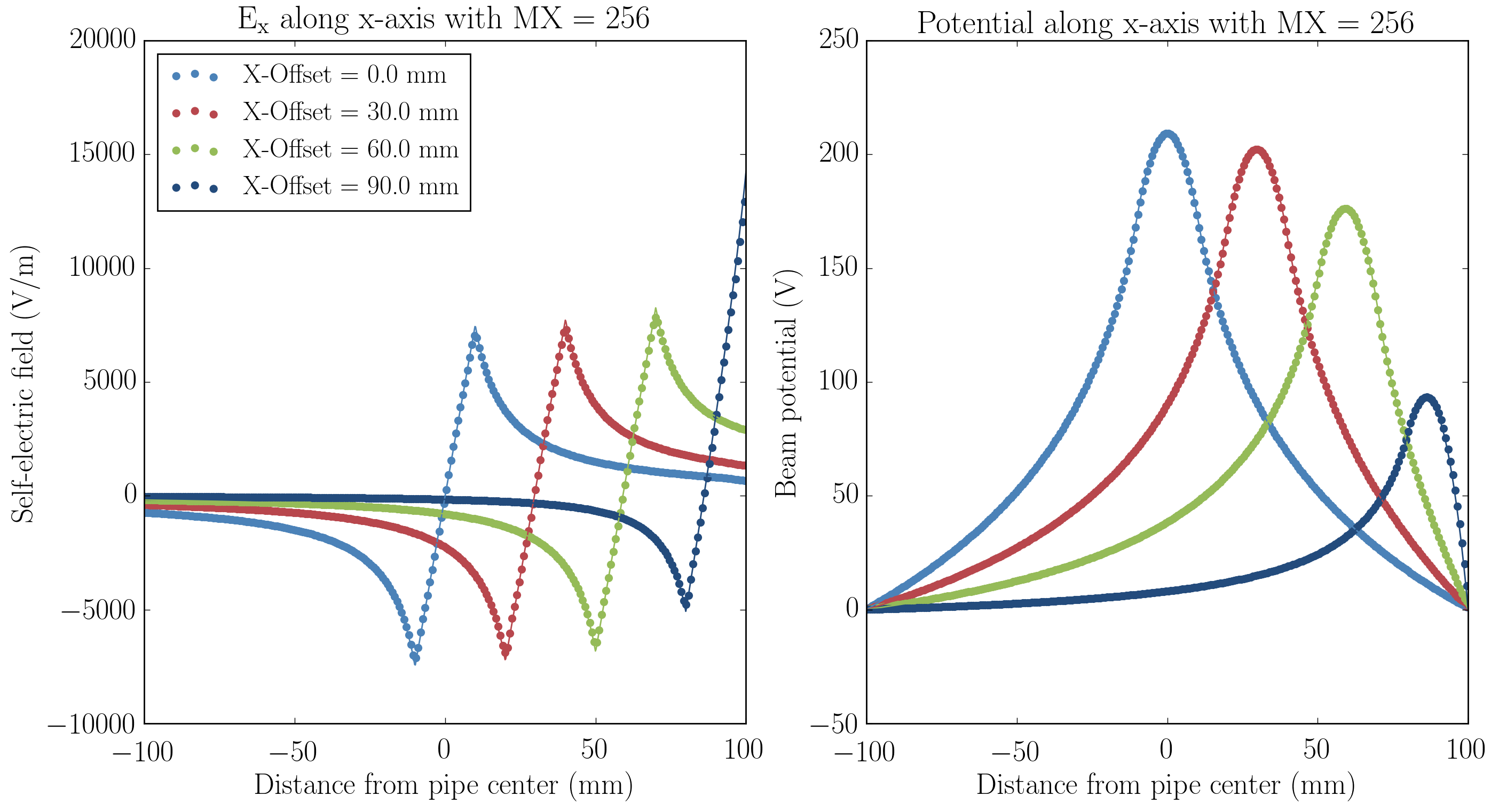}
	\caption{$\Phi$ and $E_x$ along the x-axis for different offsets $\xi$. 
	         Dots are values calculated by OPAL, while solid lines are the 
	         analytical solution. Excellent agreement can be seen with a reduced 
	         $0.01 < \chi^2_{\mathrm{red}} < 0.02$ for the potential and 
	         $0.03 < \chi^2_{\mathrm{red}} < 1.3$ for the field.}
	\label{fig:offset_variation}
\end{figure*}

For such an infinitely long beam, $E$ and $\Phi$ are independent of z,
$E_{z}(x,y) = 0$, and $E_{x,y}(x,y)$ and $\Phi(x,y)$ can be calculated from
Poisson's equation using the method of image charges.
With
\BE
\label{eqn:deltaphi_noneut}
\Delta\phi = \frac{I}{4 \pi \epsilon_0 v_b}.
\EE
(where $I$ is the beam current and $v_b$ the beam velocity),
the resulting expressions for inside and outside
(superscript ``in'' and ``out'') of the beam 
envelope are then:
\begin{align}
\Phi^{\mathrm{in}} &=
\Delta\phi\left[1 + \ln\left(\frac{\chi}{r_p^2 r_b^2}\right)
- \frac{(x-\xi)^2 + y^2}{r_b^2}\right] \label{eqn:epot_start}  \\
\Phi^{\mathrm{out}} &= 
\Delta \phi\ln \left[ \frac{\chi}{r_p^2(x-\xi)^2 + r_p^2 y^2}\right] \\
E_{x}^{\mathrm{in}} &= 
2\Delta\phi\left[ \frac{x-\xi}{r_b^2} - \eta_x\right] \\
E_{y}^{\mathrm{in}} &=
2\Delta\phi\left[ \frac{y}{r_b^2} - \eta_y\right] \\
E_{x}^{\mathrm{out}} &= 
2\Delta\phi\left[\frac{x-\xi}{(x-\xi)^2+y^2} - \eta_x\right] \\
E_{y}^{\mathrm{out}} &=
2\Delta\phi\left[\frac{y}{(x-\xi)^2+y^2} - \eta_y\right]\label{eqn:epot_end}
\end{align}
where
\begin{eqnarray*}
\eta_x & = & \frac{\xi^2 x - \xi r_p^2}{\chi}\\
\eta_y & = & \frac{\xi^2y}{\chi}
\end{eqnarray*}
and 
\begin{equation*}
\chi = \xi^2y^2+(\xi x - r_p^2)^2.
\end{equation*}

A wide parameter space was mapped in terms of mesh size, number of particles,
beam length, and position of the beam inside the beam pipe. As can be expected,
the comparison between theory and simulation gets better with higher resolution 
(i.e. higher number of mesh cells), and larger number of particles. 
The reduced $\chi$-square was chosen to compare the 
simulated results to the theoretical prediction and a plot of 
$\chi^2_{\mathrm{red}}$ for variation of mesh size and particle number 
is shown in \figref{fig:chisquares}. It can be seen that the agreement is 
better for a centered beam and so it is especially important to choose a high 
enough resolution when the beam is close to the beam pipe (or other electrodes
in the system). For this particular case, it was found that a total number
of mesh cells of 256 in x-direction ($\approx 25$ across the beam diameter) 
together with $\approx 2\cdot10^6$ particles gave excellent agreement
even when the beam was touching the pipe, with only slight or no further 
improvement at larger numbers.

As another representative example, the OPAL results of a 0.6 m long beam in a 10 cm diameter
beam pipe, using 2048000 particles and a mesh of dimensions 256 x 128 x 512, are
plotted together with the analytical solution from equations
\ref{eqn:epot_start} -- \ref{eqn:epot_end} for a beam with varying offset $\xi$ in
x-direction from the center of the beam pipe in \figref{fig:offset_variation}.

In summary, the SAAMG solver performed as expected when tested with the simple
test-case of a quasi-infinite uniform beam in a conducting pipe. In the next section,
the solver will be applied to a real world problem and results will be compared to 
measurements.

\section{Bench-marking Against Experiments\label{sec:BCS}}
An important step in bench-marking new simulation software is the comparison 
with experiments. During the summers of 2013 and 2014, a measurement campaign 
was held at Best Cyclotron Systems Inc. (BCS), to test the production of
a high intensity \htp beam in an off-resonance ECR ion source and its injection 
into a compact (test) cyclotron through a spiral inflector. 
These tests were performed within the ongoing R\&D effort for the \DD and 
IsoDAR experiments (cf. next section) and provided a good opportunity to 
compare results of injected beam measurements with OPAL simulations using the 
new spiral inflector capability.

\subsection{\DD and IsoDAR}
\begin{figure}[!t]
\centering
	\includegraphics[width=0.8\columnwidth]
	                {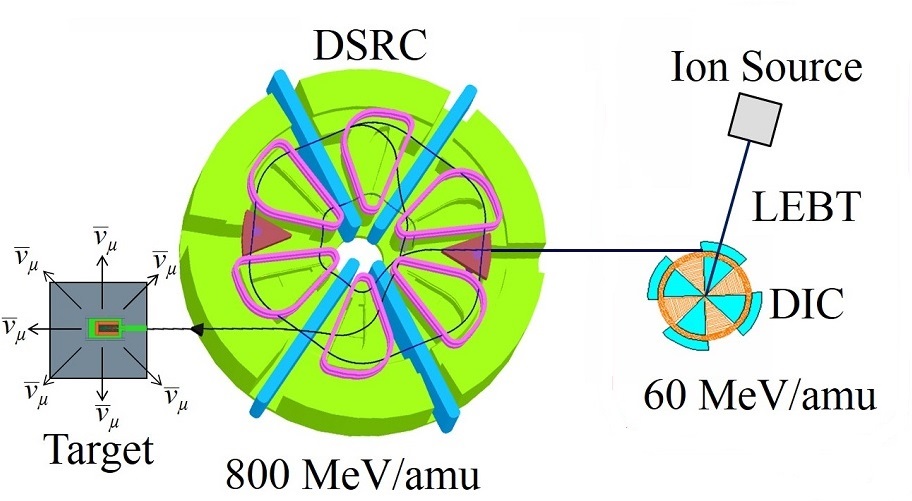}
	\caption{Schematic of the \DD facility. \htp is produced in the ion source,
	         transported to the \DD Injector Cyclotrons (DIC) and accelerated 
	         to \SI{60}{MeV/amu}. Ions are subsequently extracted from the cyclotron
	         and injected into the \DD Superconducting Ring Cyclotron (DSRC),
	         where they are accelerated to \SI{800}{MeV/amu}. During the highly 
	         efficient stripping extraction, \SI{5}{emA} of \htp becomes \SI{10}{emA} of
	         protons which impinge on the neutrino production target.}
	\label{fig:daedalus_cartoon}
\end{figure}

The Decay At-rest Experiment for $\delta_{\mathrm{CP}}$ studies At a 
Laboratory for Underground Science (\DD) \cite{abs:daedalus, aberle:daedalus} 
is a proposed experiment to measure CP violation in the neutrino sector.
A schematic view of one \DD complex is shown in \figref{fig:daedalus_cartoon}. 
\htp is produced in an ion source, transported to the \DD Injector Cyclotrons 
(DIC), and accelerated to 60 MeV/amu. The reason for using \htp instead of
protons is to overcome space charge limitations of the high required beam
intensity of 10 emA of protons on target. \htp gives 2 protons for each 
unit of charge transported, thus mitigating the risk.
The ions are subsequently extracted from the cyclotron and injected into the 
\DD Superconducting Ring Cyclotron (DSRC) where they are accelerated to 
800 MeV/amu. During the highly efficient stripping extraction, the 5 emA of 
\htp become 10 emA of protons which impinge on the neutrino production target
(carbon) producing a neutrino beam virtually devoid of \nuebar.
In a large detector, one can then look for \nuebar appearance through 
neutrino oscillations.
As is depicted in \figref{fig:daedalus_cartoon}, the injector stage of \DD
can be used for another experiment: 
The Isotope Decay At Rest experiment IsoDAR \cite{adelmann:isodar, bungau:isodar}.
In IsoDAR, the \SI{60}{MeV/amu} \htp will impinge on a beryllium target creating 
a high neutron flux. The neutrons are captured on $^7$Li surrounding the target.
The resulting $^8$Li beta-decays producing a very pure, isotropic \nuebar beam
which can be used for \nuebar disappearance experiments. IsoDAR is a definitive
search for so-called ``sterile neutrinos'', proposed new fundamental particles
that could explain anomalies seen in previous neutrino oscillation experiments.

At the moment, \opalcycl is used for the simulation of three very important parts 
of the \DD and IsoDAR systems:
\begin{enumerate}
\itemsep1pt
\item The spiral inflector
\item The \DD Injector Cyclotron (DIC), which is identical to the 
      IsoDAR cyclotron. 
\item The \DD Superconducting Ring Cyclotron (DSRC) for final acceleration. 
\end{enumerate}
For the topic of bench-marking, we will restrict ourselves to item 1., the injection
through the spiral inflector.

\subsection{The Teststand}
As mentioned before, the results of the injection tests are 
reported in detail in \cite{winklehner:bcs_tests}. Here, we will
summarize the items pertinent to a comparison to \opal, 
specifically, how we obtain the particle distribution at the
end of the LEBT (entrance of the cyclotron), used as initial 
beam in the subsequent injection simulations with the SAAMG
solver.

The test stand was comprised of the following parts:
\begin{enumerate}
\itemsep1pt
	\item Versatile Ion Source (VIS) \cite{miracoli:vis1}. An
	      off-resonance Electron Cyclotron Resonance (ECR) ion
	      source.
	\item The Low Energy Beam Transport (LEBT). The LEBT
	      contained:
	      \begin{enumerate}
	      \itemsep1pt
	      \item First solenoid magnet, for separation of
	      protons and \htp.
	      \item Beam diagnostics (emittance scanner,
	      faraday cup).
	      \item Steering magnets.
	      \item Second solenoid magnet (final focusing into the
	      cyclotron).
	      \end{enumerate}
	\item Cyclotron with spiral inflector.
\end{enumerate}

During the experiment, it was possible to transport up to 8 mA 
of \htp as a DC beam along the LEBT to the cyclotron and 
transfer 95\% of it through the spiral inflector onto a 
paddle probe. The 4-rms normalized emittances stayed below 
1.25 $\pi$-mm-mrad. Capture efficiency into the RF ``bucket''
was 1-2\% because of reduced dee voltage 
($\mathrm{V}_{\textrm{dee}}$) due to an under-performing
RF amplifier (cf. discussion in \cite{winklehner:bcs_tests}).

\subsection{Initial Conditions}

\begin{figure}[!t]
\centering
	\includegraphics[width=1.0\columnwidth]
	                {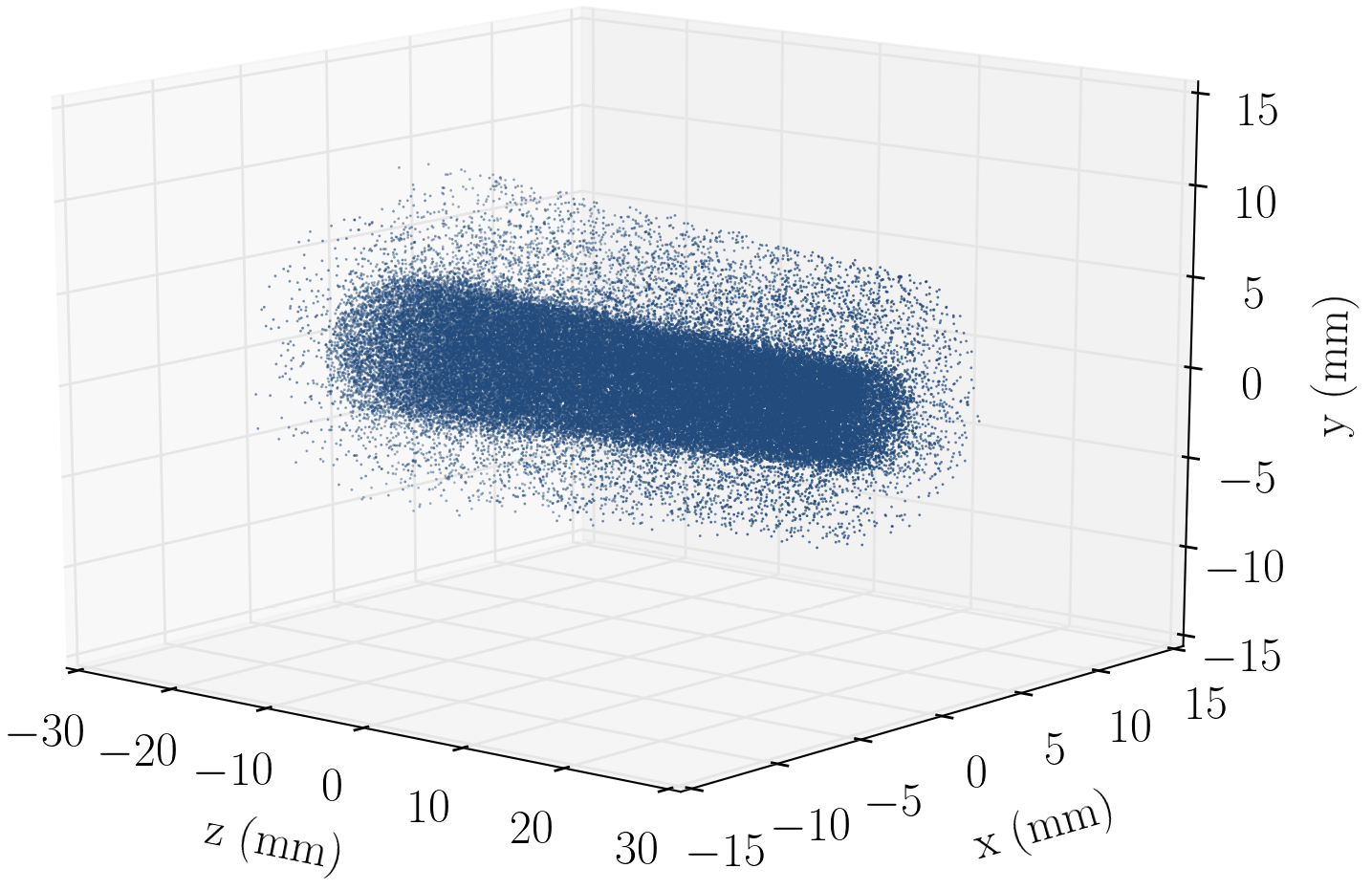}
	\caption{Initial distribution for injection through
	         the spiral inflector, obtained by carefully
	         simulating the LEBT \cite{winklehner:bcs_tests}.
	         The length of the bunch corresponds to one
	         full RF period at \SI{49.2}{MHz} and injection energy 
	         of \SI{62.7}{keV}, centered at the synchronous phase, 
	         thus, to first order, representing a DC beam.}
	\label{fig:init_dist}
\end{figure}

The quality of any simulation result depends on the initial 
conditions. In the case of the \opal SAAMG simulations of
the injection through the spiral inflector, the
initial particle distribution consisted of 66021 particles
obtained by carefully simulating the ion source extraction
(using KOBRA-INP \cite{spaedke:kobra}) and the subsequent
LEBT (using WARP \cite{grote:warp1}) and comparing 
the simulation results to the measurements, with good agreement 
as reported in \cite{winklehner:bcs_tests}. 
During the WARP simulations
of the LEBT, the ``xy-slice-mode'' was used in which
the self-fields are calculated only for the
transverse direction (assuming only very slow changes in 
beam diameter along the z-axis compared to the length of each
simulation step) and neglecting longitudinal self-fields 
(which is a sensible approach for DC beams).
Space charge compensation played a big role in order
to obtain good agreement and was taken into account 
using a semi-analytical formula \cite{winklehner:scc3}.
The final particle distribution that was obtained for the 
set of parameters recorded during the measurements is shown
in \figref{fig:init_dist}.
It should be noted that the bunch was generated from the 
xy-slice at a position 13 cm away from the cyclotron mid-plane
and coaxial with the cyclotron center by randomly backward 
and forward-projecting particles according to their respective momenta. 
It can be seen that this beam enters the 
spiral inflector converging, which has been found 
experimentally to give the best injection efficiency.
The important parameters of the injected beam are listed
in \tabref{tab:init_dist}.

\begin{table}[!b]
	\caption{Beam and cyclotron parameters for inflection and acceleration studies.}
	\label{tab:init_dist}
	\centering
    \vspace{5pt}
    \renewcommand{\arraystretch}{1.25}
		\begin{tabular}{ll}
            \hline
            \textbf{Parameter} & \textbf{Value} \\
            \hline \hline
            Species & \htp \\
            Initial Beam Energy & 62.7 keV \\
            2-rms Beam Diameter & 10.6 mm \\
            4-rms Normalized Emittance & 1.19 $\pi$-mm-mrad \\
            Cyclotron Magnetic Field & 1.1 T average\\
            Spiral Inflector Upper / Lower & -10.0 kV/+10.15 kV \\
            Beam Current & 7.0 mA \\
            Approximate Dee Voltage & $\approx60$ kV \\
            \hline
		\end{tabular}
\end{table}

\subsection{Results}
\begin{figure*}[tpbh]
	\centering
		\includegraphics[width=0.75\textwidth]
	                    {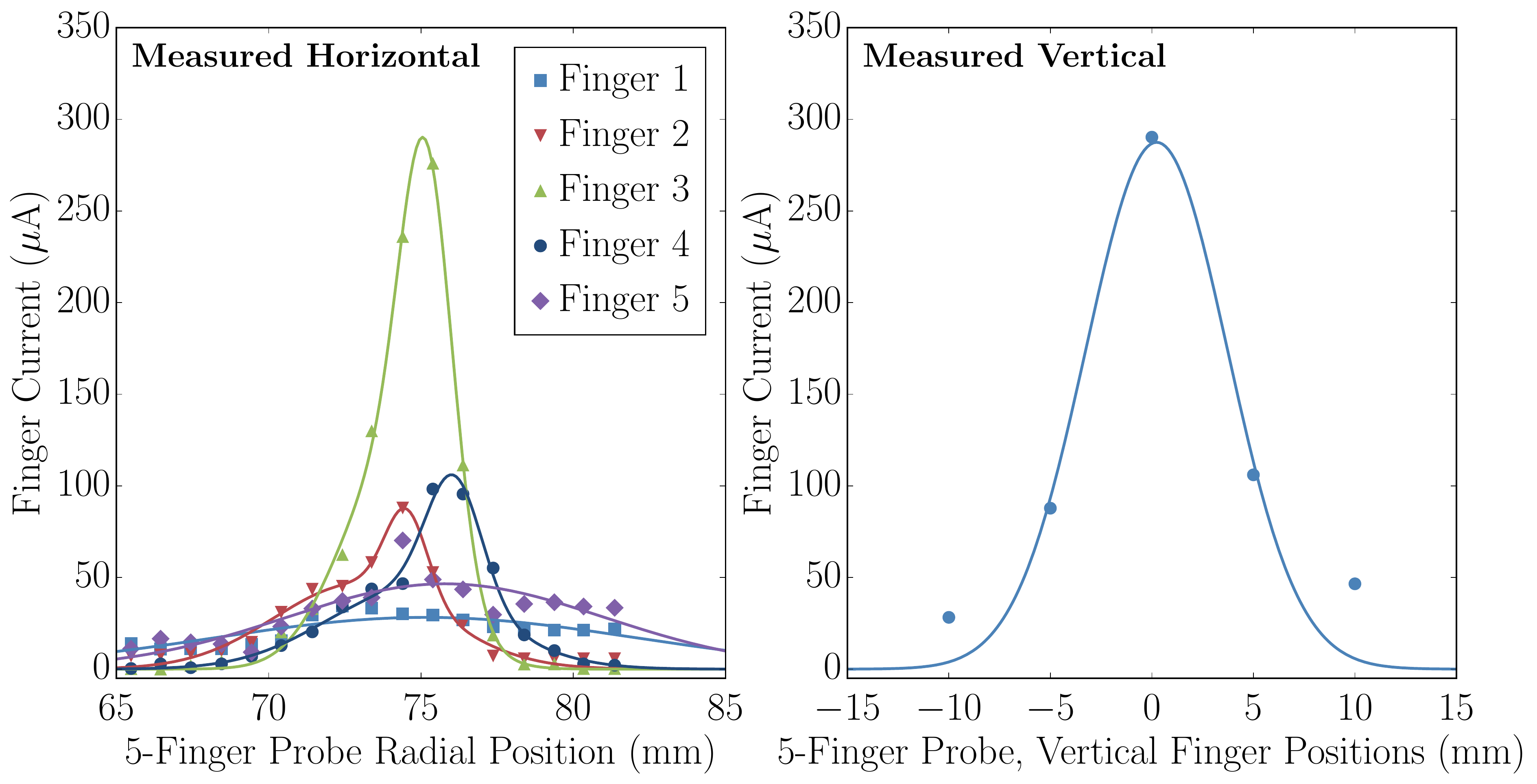}
	    \includegraphics[width=0.75\textwidth]
	                    {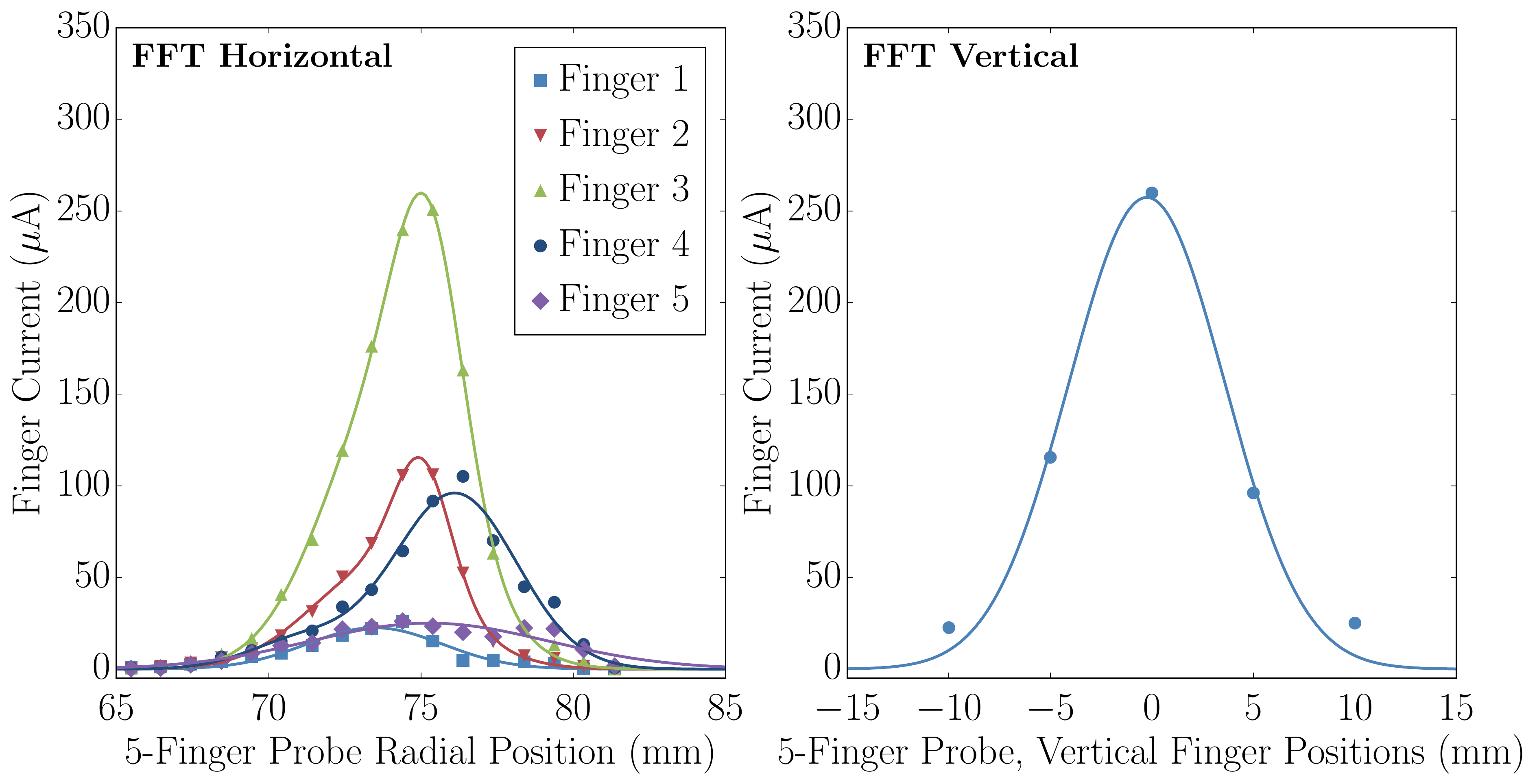}
	    \includegraphics[width=0.75\textwidth]
	                    {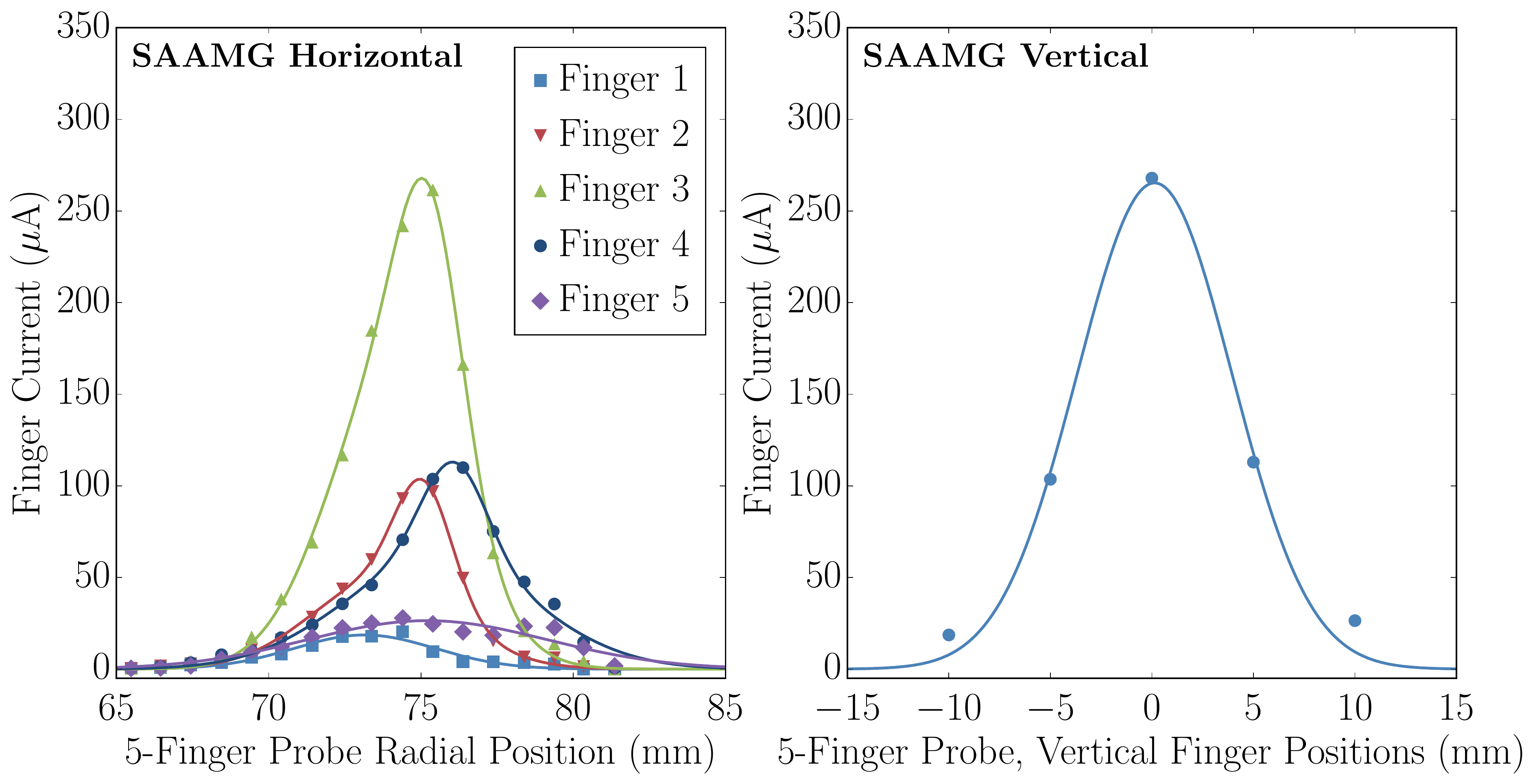}
	\caption{Measurement and simulation of a 5-finger probe (for description cf. to text)
	        in the first turn of the BCS test cyclotron central region, 45\degree
	        after the exit of the spiral inflector. Each finger has a main peak and a
	        tail towards lower radius. These are the particles that are not sufficiently 
	        accelerated. To guide the eye and to compare with the simulations, 
	        the data for each finger is fitted with two Gaussians.}
	\label{fig:bcs_5finger}
\end{figure*}

The beam described in the previous section was then transported
through the spiral inflector using the standard FFT and the new 
SAAMG field solver described in Section \ref{ssec:fieldsolver}, 
and putting in place the geometry seen in \figref{fig:geometry}.
\begin{figure}[!t]
\centering
	\includegraphics[width=0.78\columnwidth]
	                {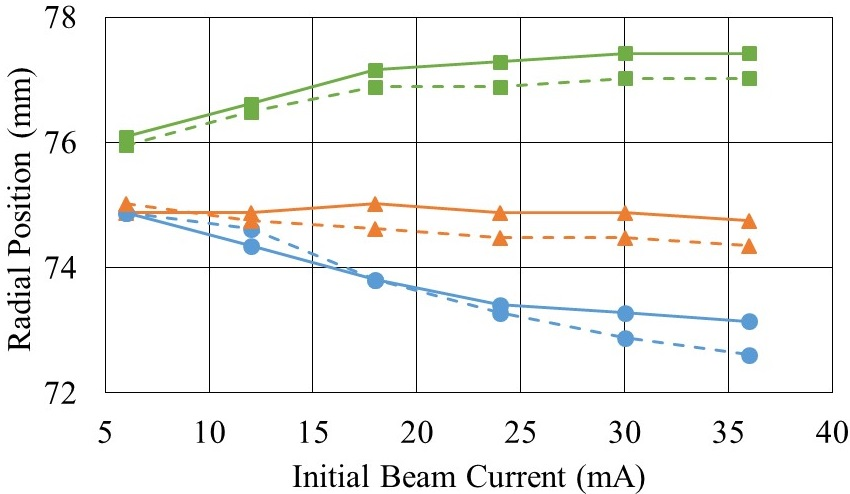}
	\caption{Radial positions of the peak of each finger of the 
	         five-finger probe (cf. text). Blue (circles): Finger 2, 
	         orange (triangles): Finger 3, green (squares): Finger 4,
	         solid: FFT, dashed: SAAMG.}
	\label{fig:higher_currents}
\end{figure}
Inside the cyclotron, 45\degree after the exit of the spiral inflector,
a radial probe was placed which had 5 fingers of $\approx5$ mm vertical 
extent, and $\approx1$ mm radial extent each. On these fingers, the 
electrical beam current was measured. The probe was slowly
moved from a position blocking the beam completely, to just outside 
of the radial extent of first turn, thereby giving the beam
current distribution shown in the top plot of \figref{fig:bcs_5finger}.
In the same plot, the results from OPAL simulations using the same 
parameters as recorded during the measurement (see Table \ref{tab:init_dist})
are plotted. Good qualitative agreement can be seen for both the FFT and the 
SAAMG solver. Due to the tail towards low radius, two Gaussians are used 
to fit each peak in the left column. The full widths at half maximum (FWHM) 
of the dominant Gaussians are listed in Table \ref{tab:fwhm}.
There seems to be a slight shift towards higher vertical position that 
is better reproduced in the SAAMG solver, but this is well within the
systematic errors of the measurement and how well the initial parameters
like magnetic field, spiral inflector voltage and beam distribution
were known, hence the conclusion is that both FFT and SAAMG reproduce 
the measured radial--vertical beam distribution equally well for a 
6 mA beam. This shows that the SAAMG solver is working as expected. 

\begin{table}[!b]
	\caption{Full width at half maximum (FWHM) for the measured distribution and
	         the OPAL results using the FFT and the SAAMG solver.}
	\label{tab:fwhm}
	\centering
    \vspace{5pt}
    \renewcommand{\arraystretch}{1.25}
		\begin{tabular}{lrrr}
            \hline
            & \textbf{Measured} & \textbf{FFT} & \textbf{SAAMG}\\
            \hline \hline
			\textbf{Finger 1} & 16.26 mm & 5.21 mm &  5.47 mm\\
			\textbf{Finger 2} & 1.64 mm & 2.19 mm & 2.19 mm\\
			\textbf{Finger 3} & 2.03 mm & 2.61 mm & 2.52 mm\\
			\textbf{Finger 4} & 2.15 mm & 4.75 mm & 2.50 mm\\
			\textbf{Finger 5} & 12.31 mm & 9.26 mm & 9.28 mm\\
			\hline
			\textbf{Vertical} & 8.22 mm & 9.05 mm & 8.99 mm\\
            \hline
		\end{tabular}
\end{table}

\subsubsection{Higher beam currents}

For initial beam currents up to 36 mA, the results of the FFT 
and SAAMG solvers start to show stronger (but still fairly subtle) 
discrepancies which can be attributed to the image charges on the 
electrodes only included with the SAAMG solver. 
An example is shown in \figref{fig:higher_currents} where
the expected spreading of the peak positions is accompanied by a 
noticeable overall shift towards smaller radii in case of the SAAMG
solver.



\section{Conclusion\label{sec:conclusion}}

\subsection{Summary and Discussion}
For the first time a comprehensive and precise beam 
dynamics simulation model, from the ion source, throughout  
the LEBT, into the central region of a cyclotron was presented. 
The central region includes the spiral inflector and the
first accelerating gap.
From the exit of the LEBT, the open 
source code \opal\ was used for the beam transport through
the spiral inflector and the first turn of the test cyclotron.
Key ingredients of the model are the the flexible handling of 
the complex geometry, and the field solvers for space charge 
calculation. In comparison with first measurements, both the FFT 
and the SAAMG solver perform well, with hints that image charge 
effects become more important at higher currents, where use of the
SAAMG solver allows including the complicated boundary conditions
posed by the electrode system. 
These ingredients - geometry and field solvers (FFT and SAAMG) 
are now included with \opal.
Validation of the model included simple test cases and 
comparison to measurements from a dedicated cyclotron test stand, 
injecting a DC beam of 7 mA of {\htp}. Both yielded good agreement.
The level of detail available in this model now 
allows us to obtain a detailed understanding, and predict 
the complicated beam dynamics in the high current 
compact IsoDAR cyclotron.

\subsection{Outlook}
Given this benchmarked model, a full start to end 
simulation of the IsoDAR cyclotron is ongoing, using first 
the detailed geometry for acceleration up to 1 MeV/amu,
and then a simplified model with FFT only for the subsequent acceleration
up to \SI{60}{MeV/amu}.
Recently, a proposal was put forward to test direct 
injection into the compact IsoDAR cyclotron
using a Radio Frequency Quadrupole (RFQ) \cite{winklehner:rfq2}. 
For the design of this device, \opalcycl and
the new \verb|SPIRAL| mode will play an essential role.

\begin{acknowledgments}

This work was supported by the US National Science Foundation under award \#1505858
and the corresponding author was partly supported by the Bose Foundation.
The research at PSI leading to these results has received funding from the European Community's 
Seventh Framework Programme (FP7/2007-2013) under grant agreement \#290605 (PSI-FELLOW/COFUND).
Furthermore, the authors would like to express their gratitude to Best Cyclotron Systems, Inc.
in Vancouver, for hosting the 1 MeV cyclotron injection tests, 
and the INFN-LNS ion source group in Catania, for the loan of the VIS.

\end{acknowledgments}

\bibliography{SI_Bibliography}

\end{document}
%